\documentclass{llncs}
\usepackage{caption}
\usepackage{xspace}
\usepackage{algorithm}
\usepackage{algorithmic}
\usepackage{graphicx}
\usepackage{url}
\usepackage{multirow}
\usepackage{wrapfig}
\usepackage{subfig}

\usepackage{amsmath}
\usepackage{amssymb}

\usepackage[margins]{trackchanges} 

\newsubfloat[position=top, listofformat=subsimple]{algorithm}

\newcommand{\secspace}{\vspace{-2mm}}
\newcommand{\subsecspacea}{\vspace{-4mm}}
\newcommand{\subsecspaceb}{\vspace{-1mm}}

\addeditor{OS} \addeditor{BR} \addeditor{MH} \addeditor{DH} 

\newcommand{\boost}{\textsc{Boost}\xspace}
\newcommand{\cgal}{\textsc{Cgal}\xspace}
\newcommand{\core}{\textsc{Core}\xspace}

\newcommand{\CC}{C\raise.08ex\hbox{\tt ++}\xspace}

\newcommand{\Cfree}{\ensuremath{\calC_{\rm free}}\xspace}
\newcommand{\Cforb}{\ensuremath{\calC_{\rm forb}}\xspace}
\newcommand{\MMS}{MMS\xspace}
\newcommand{\FSC}{\ensuremath{F\!\!\:S\!\!\:C}\xspace}
\newcommand{\FSCs}{\ensuremath{F\!\!\:S\!\!\:Cs}\xspace}
\newcommand{\DOF}{\textit{dof}\xspace}
\newcommand{\DOFs}{\textit{dofs}\xspace}

\newcommand{\R}{\ensuremath{\mathbb{R}}}
\newcommand{\Z}{\ensuremath{\mathbb{Z}}}

\newcommand{\N}{\ensuremath{\mathbb{N}}}

\newcommand{\etal}{{et~al.}\xspace}

\def\naive{{na\"{\i}ve}\xspace}

\newcommand{\calC}{\ensuremath{\mathcal{C}}\xspace}

\newcommand{\calG}{\ensuremath{\mathcal{G}}\xspace}

  \renewenvironment{thebibliography}[1]{%
    \begin{oldthebibliography}{#1}%
      \footnotesize
      \setlength{\parskip}{0ex}%
      \setlength{\itemsep}{0ex}%
  }%
  {%
    \end{oldthebibliography}%
  }

\newcommand{\ignore}[1]{}
\newcommand{\first}[2]{#1}

\def\marrow{\marginpar[\hfill$\longrightarrow$]{$\longleftarrow$}}

\first{
\def\dan#1{\textcolor{red}{\textsc{Danny says: }{\marrow\sf #1}}}
\def\michael#1{\textcolor{blue}{\textsc{Michael says: }{\marrow\sf #1}}}
\def\barak#1{\textcolor{cyan}{\textsc{Barak says: }{\marrow\sf #1}}}
\def\oren#1{\textcolor{magenta}{\textsc{Oren says: }{\marrow\sf #1}}}
}{
\newcommand{\dan}[1]{}
\newcommand{\michael}[1]{}
\newcommand{\barak}[1]{}
\newcommand{\oren}[1]{}
}

\newcommand{\niceX}{$\times$}
\newcommand{\niceV}{\checkmark}

\begin{document}


\mainmatter

\title{Motion Planning via Manifold Samples
\thanks{This work has been supported in part by the 7th Framework
Programme for Research of the European Commission, under
FET-Open grant number 255827 (CGL---Computational Geometry
Learning), by the German-Israeli Foundation (grant no.
969/07), and by the Hermann Minkowski--Minerva Center for
Geometry at Tel Aviv University.}
}
\author{Oren Salzman\inst{1} \and Michael Hemmer\inst{1}\and Barak Raveh\inst{1,2} \and Dan Halperin\inst{1}}

\tocauthor{%
Oren Salzman (Tel-Aviv University)
Michael Hemmer (Tel-Aviv University) ,
Barak Raveh (Tel-Aviv University, Hebrew University),
Dan Halperin (Tel-Aviv University)}

\institute{
Tel-Aviv University, Israel \\
\and
Hebrew University, Israel
}

\maketitle

\begin{abstract}
We present a general and modular algorithmic framework for path planning of robots. Our framework combines  geometric methods for exact and complete analysis of low-dimensional configuration spaces, together with practical, considerably simpler sampling-based approaches that are appropriate for higher dimensions. In order to facilitate the transfer of advanced geometric algorithms into practical use, we suggest taking samples that are \emph{entire low-dimensional manifolds of the configuration space} that capture the connectivity of the configuration space much better than isolated point samples. Geometric algorithms for analysis of low-dimensional manifolds then provide powerful primitive operations.
The modular design of the framework enables independent optimization of each modular component. Indeed, we have developed, implemented and optimized a primitive operation for complete and exact combinatorial analysis of a certain set of manifolds, using arrangements of curves of rational functions and concepts of generic programming. This in turn enabled us to implement our framework for the concrete case of a polygonal robot translating and rotating amidst polygonal obstacles. We demonstrate that the integration of several carefully engineered components leads to 
significant speedup over the popular PRM sampling-based algorithm, which represents the more simplistic approach that is prevalent in practice. We foresee possible extensions of our framework to solving high-dimensional problems beyond motion planning.

\end{abstract}


\section{Introduction}
\secspace
\label{sec:introduction}
Motion planning is a fundamental research topic in robotics with
applications in diverse domains such as graphical animation,
surgical planning, computational biology and computer games. For a
general overview of the subject and its applications 
see~\cite{planning-survey-choset,Latombe-Robot-Mot-Plan,planning-survey-lavalle}. 
In its basic
form, the motion-planning problem is to find a collision-free path
for a robot or a moving object $R$ in a \emph{workspace} cluttered
with static obstacles. The spatial pose of $R$, or the
\emph{configuration} of $R$, is uniquely defined by some set of
parameters, the degrees of freedom (\DOF{s}) of $R$. The set of all
robot configurations \calC is termed the \emph{configuration space}
of the robot, and decomposes into the disjoint sets of free and
forbidden configurations, namely  $\Cfree$ and $\Cforb$,
respectively. Thus, it is common to rephrase the motion-planning
problem as the problem of moving $R$ from a start configuration
$q_s$ to a target configuration $q_t$ in a path that is fully
contained within $\Cfree$.

\vspace{1mm}
\textbf{Analytic solutions to the general motion planning problem:}
The motion-planning problem is computationally hard with respect to
the number of~\DOFs~\cite{Piano-complexity}, yet much research has
been devoted to solving the general problem and its various
instances using geometric, algebraic and
combinatorial tools. 
The configuration-space formalism was introduced by 
Lozano-Perez~\cite{Lozano-Perez80spatialplanning} in the early 1980's. 
Schwartz and Sharir proposed the first general
algorithm for solving the motion planning problem, with running time
that is doubly-exponential in the number of \DOF{s}~\cite{PianoII}.
Singly exponential-time algorithms have followed~\cite{Basu03,Canny-complexity,Chazelle199177}, but
are generally considered too complicated to be implemented in
practice.

\vspace{1mm}
\textbf{Solutions to low-dimensional instances of the problem:}
Although the general motion-planning problem cannot be 
efficiently solved analytically, more efficient algorithms 
have been proposed for various low-dimensional instances~\cite{Latombe-Robot-Mot-Plan}, 
such as translating a
polygonal or polyhedral robot~\cite{AronovS97,Lozano-Perez80spatialplanning}, 
and translation with rotation of a polygonal robot in the plane~\cite{AvnaimBF88,HS96,PianoI}.
For a survey of related approaches see~\cite{planning-survey-sharir}. 
Moreover, considerable advances in robust implementation of 
computational geometry algorithms in recent years  have led 
to a set of implemented tools that are of interest in this context. 
Minkowski sums, which allow representation of the configuration 
space of a translating robot, have robust and exact planar and 
3-dimensional implementations~\cite{fogel-mink,Hachenberger09,w-eecpm-06}. 
Likewise, implementations of planar 
arrangements\footnote{A subdivision of the plane into zero-dimensional, 
one-dimensional and two-dimensional cells, called vertices, edges and 
faces, respectively induced by the curves.} 
for curves ~\cite[C.30]{cgal}, 
are essential components 
in~\cite{PianoII}.

\vspace{1mm}
\textbf{Sampling-based approaches to motion planning:}
The sampling-based approach to motion-planning
has extended the applicability of motion planning algorithms
beyond the restricted subset of problems that can be solved efficiently by
exact algorithms~\cite{planning-survey-choset,planning-survey-lavalle}. 
Sampling-based motion planning algorithms,
such as 
Probabilistic Roadmaps (PRM)~\cite{Kavraki-PRM}, Expansive Space Trees 
(EST)~\cite{Hsu99} and Rapidly-exploring
Random Trees (RRT)~\cite{Lavalle-RRT}, 
 as well as their many variants,
 aim to capture the connectivity of $\Cfree$ in a graph data structure, via
 random sampling of robot configurations. 
 This can be done either in a multi-query setting,
to efficiently answer multiple queries for the same scenario, as in the 
PRM algorithm,
or in a single-query setting, as in the RRT and EST algorithms.
For a general survey on the field see~\cite{planning-survey-choset}.
Importantly, the PRM
and RRT
algorithms
were both shown to be probabilistically 
complete~\cite{Kavraki98,Kuffner00,LK02}, that is, they are
guaranteed to find a valid solution, if one exists. However,
 the required running time for finding such a solution cannot be
computed for new queries at run-time, and the proper usage
of sampling-based approaches may still be considered somewhat of an art.
Moreover, sampling-based methods are also considered sensitive to tight passages
in the configuration space, due to the high-probability of missing the passage.

\vspace{1mm}
\textbf{Hybrid methods for motion-planning:}
Few hybrid methods attempt to combine both deterministic 
and probabilistic planning strategies. Hirsch and 
Halperin~\cite{hybrid-disks} studied two-disc motion 
planning by exactly decomposing the configuration space 
of each robot, then combining the two solutions to a set 
of free, forbidden and mixed cells, and using PRM to 
construct the final connectivity graph.  
Zhang et al.~\cite{ZhangKM07} used PRM in conjunction 
with approximate cell decomposition, which also divides 
space to free, forbidden and mixed cells. Other studies
 have suggested to connect a dense set of near-by 
configuration space ``slices''. Each slice is decomposed 
 to free and forbidden cells, but adjacent slices 
are connected in an inexact manner, by e.g., identifying 
overlaps between adjacent slices~\cite[pp.~283-287]{CG-alg-app}, 
or heuristic interpolation and local-planning~\cite{hybrid-mink}.  
In~\cite{YangS06} a 6 \DOF RRT planner is presented with 
a 3~\DOF local planner hybridizing probabilistic, 
heuristic and deterministic methods.

\subsecspacea
\subsection{Contribution}
\subsecspaceb
In this study, we present a novel general scheme for motion
planing via manifold samples (\MMS), which extends sampling-based techniques like PRM as
follows: Instead of sampling isolated robot configurations, we sample
\emph{entire low-dimensional manifolds}, which can be analyzed by
complete and exact methods for decomposing space. This yields an explicit representation of
maximal connected patches of free configurations on each manifold,
and provides a much better coverage of the configuration space
compared to isolated point samples. At the same time, the manifold
samples are deliberately chosen such that they are likely to
intersect each other, which allows to establish connections among
different manifolds. The general scheme of \MMS is illustrated in Figure
\ref{fig:conf_space_slicing}. A detailed discussion of the scheme is presented in
Section~\ref{sec:gen_scheme}. 

In Section~\ref{sec:alg_framework}, we discuss the application of \MMS to the concrete case
of a polygonal robot translating and rotating in the plane
amidst polygonal obstacles. We present in detail
appropriate families of manifolds as well as filtering schemes that
should also be of interest for other scenarios.
Although our software is prototypical, we 
emphasize that the achieved results are due to careful design and
implementation on all levels.  
In particular, in Section~\ref{sec:impl_details} we present 
an exact analytic solution and efficient implementation to a 
motion planning problem instance: moving a polygonal robot 
in the plane with rotation and translation 
\emph{along an arbitrary axis}. To the best of our knowledge
 the problem has not been analytically studied before. 
The implementation involves advanced algebraic and 
extension of state-of-the-art applied geometry tools. 
In Section~\ref{sec:exp_results} we
present experimental results, which show our method's superior behavior for
several test cases vis-\`{a}-vis a common implementation of the
sampling-based PRM algorithm. 
For example, in a tight passage scenario we demonstrate a 27-fold improvement.  
 We conclude with a discussion of
extensions of our scheme, which we anticipate could greatly widen
the scope of applicability of sampling-based methods for motion planning by
combining them with strong analytic tools in a 
straightforward manner.
\section{General Scheme for Planning with Manifold Samples}
\label{sec:gen_scheme}
\begin{figure}
    \vspace{-7mm}
  \centering
   \begin{center}
       \includegraphics[width=0.9\textwidth]{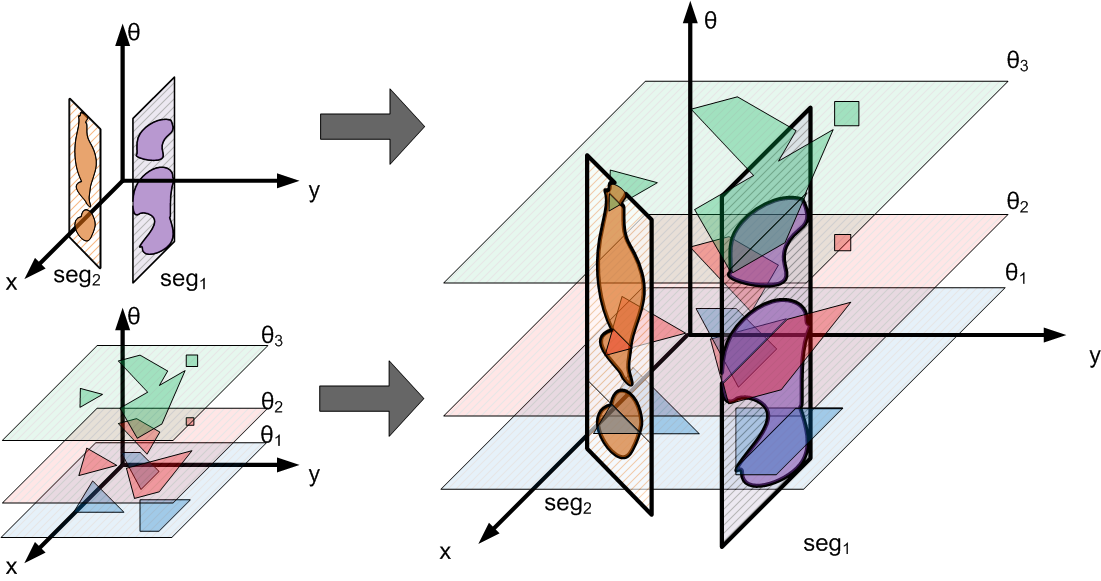}
   \end{center}
   \vspace{-4mm}
   \caption{Three-dimensional configuration space:
   					The left side illustrates two families of manifolds where the decomposed cells 
   					are darkly shaded. The right side illustrates their intersection that induces 
   					the graph $\calG$.}
   \label{fig:conf_space_slicing}
   \vspace{-4mm}
\end{figure}

\secspace
\textbf{Preprocessing---Constructing connectivity graph:}
We propose a multi-query planner for motion planning problems in a
possibly high-dimensional configuration space. The preprocessing
stage constructs the \emph{connectivity graph} of $\calC$, a data structure that captures the connectivity of
$\calC$ using manifolds as samples. 
The manifolds are decomposed into cells in $\Cfree$ and $\Cforb$ in a complete and exact manner;
 we call a cell of the decomposed manifold that lies in
$\Cfree$ a \emph{free space cell} (\FSC) and refer to the connectivity graph as 
$\calG$. The \FSCs serve as nodes in $\calG$ while two nodes in $\calG$ are
connected by an edge 
if their corresponding \FSCs intersect. See  Figure \ref{fig:conf_space_slicing} for an illustration

We formalize the preprocessing stage by considering manifolds induced by a family of
constraints $\Psi$, such that $\psi \in \Psi $ defines a manifold
$m_\psi$ of the configuration space. The construction of a
manifold $m_\psi$ and its decomposition into \FSCs
are carried out via a 
$\Psi$-primitive (denoted $P_\Psi$) applied to an element $\psi \in
\Psi$. By a slight abuse of notations we refer to an \FSC both as a cell and a node in the graph.
Using this notation, Algorithm \ref{alg:preprocess}
summarizes the construction of $\calG$. In lines 3-4, a new
manifold constraint is generated and added to the collection of
manifold constraints $X$. In lines 5-7, the manifold induced by the
new constraint is decomposed by the appropriate primitive and its
\FSCs are added to $\calG$. 

\begin{algorithm}
\caption{Construct Connectivity Graph}
\label{alg:preprocess}
\begin{algorithmic}[1]
    \STATE $V \leftarrow \emptyset$, $E \leftarrow \emptyset$, $X \leftarrow \emptyset$
    \REPEAT
        \STATE $\psi \leftarrow $ generate\_constraint($V$,$E$,$X$)
        \STATE $X \leftarrow X \cup \{ \psi \}$
        \STATE $\FSC_{m_\psi} \leftarrow P_\Psi(m_\psi)$
        \STATE $V \leftarrow V \cup \{$fsc$ |\ $fsc$ \in \FSC_{m_\psi} \}$
        \STATE $E \leftarrow E \cup \{($fsc$_1,$ fsc$_2) \ |$  $\ $fsc$_1 \in V,$ fsc$_2 \in \FSC_{m_\psi}$ ,\\ 
        				\hspace{33mm}
        				fsc$_1 \cap $fsc$_2 \neq \emptyset\ $ \&  fsc$_1 \neq $fsc$_2 \}$
    \UNTIL {stopping\_condition}
    \RETURN {$G(V,E)$}
\end{algorithmic}
\end{algorithm}

\textbf{Query:}
Once the connectivity graph $\calG$ has been constructed
it can be queried for paths between two configurations $q_s$ and
$q_t$ in the following manner: A manifold that contains $q_s$
(respectively $q_t$) in one of its \FSCs is generated and
decomposed. Its \FSCs and their appropriate edges are added to $\calG$.
 We compute a path $p$ in $\calG$ between the \FSCs
that contain $q_s$ and $q_t$. A path in
$\Cfree$ may then be computed by planning a path within each \FSC in
$p$.

\subsecspacea
\subsection{Desirable Properties of Manifold Families }
\label{subsec:defining-primitives}
\subsecspaceb
Choosing the specific set of manifold families
may depend on the concrete problem at hand, as detailed in the next section. However, it seems desirable to retain
some general properties. 
First, each manifold should be simple enough
such that it is possible to decompose it into free and forbidden
cells in a computationally effcient manner.
The choice of manifold
families should also \emph{cover} the configuration space, such that
each configuration intersects at least a single manifold $m_\psi$.
In addition, local transitions between close-by configurations
should be made possible via cross-connections of several
intersecting manifolds, which we term the \emph{spanning}
property. We anticipate that these simple and intuitive
properties (perhaps subject to some fine tuning) may lead to a proof 
of probabilistic completeness of the approach.

\subsecspacea
\subsection{Exploration and Connection Strategies}
\label{subsec:applying-primitives} 
\subsecspaceb
A \naive way to generate constraints that induce manifolds is by random sampling. Primitives may be
computationally complex and should thus be applied sparingly. We suggest a
general exploration/connection scheme and additional optimization heuristics that may be used in 
concrete implementations of the proposed general scheme.
We describe strategies in general terms, providing conceptual guidelines for concrete
implementations, as demonstrated in Section~\ref{sec:alg_framework}.

\vspace{1mm}
\textbf{Exploration and connection phases:}
Generation of constraints is done in
two phases: \emph{exploration} and \emph{connection}. 
In the exploration phase
constraints are generated such that primitives will produce \FSCs
that introduce new connected components in $\Cfree$. The aim
of the exploration phase is to increase the coverage of the
configuration space as efficiently as possible. In contrast,
in the connection phase constraints are
generated such that primitives will produce \FSCs that connect
existing connected components in $\calG$. Once a constraint is generated, $\calG$ is updated
as described above.
 Finally, we note that we can alternate between exploration and connection,
namely we can decide to further explore after some connection work
has been performed.

\textbf{Region of interest (RoI):}
\label{subsec-roi} Decomposing an entire manifold $m_\psi$ by a primitive $P_\Psi$ may be
unnecessary. Patches of $m_\psi$ may intersect $\Cfree$ in highly
explored parts or connect already well-connected parts of $\calG$
while others may intersect $\Cfree$ in sparsely explored areas or
less well connected parts of $\calG$. Identifying the regions where
the manifold is of good use (depending on the phase) and
constructing $m_\psi$ only in those regions increases the
effectiveness of $P_\Psi$ while desirably biasing the samples. We
refer to a manifold patch that is relevant in a specific phase
as the \emph{Region of Interest} - RoI of the manifold.

\vspace{1mm}
\textbf{Constraint filtering }
\label{subsec-ccf} 
Let $\psi \in \Psi$ be a constraint such that
applying $P_\Psi$ to $\psi$ yields the 
 set of \FSCs on $m_\psi$. If we are
in the \emph{connection} phase, inserting the associated
nodes into $\calG$ and intersecting them with the existing \FSCs
should connect existing connected components of $\calG$. Otherwise, the
primitive's contribution is poor. We suggest applying a
filtering predicate immediately after
generating a constraint $\psi$ to check if $P_\Psi(\psi)$
\emph{may} connect existing connected components of $\calG$. If not,
 the primitive should not be constructed and $\psi$ should be
discarded.

\section{The Case of Rigid Polygonal Motion}
\label{sec:alg_framework}
\secspace
We demonstrate the scheme suggested in Section~\ref{sec:gen_scheme} by 
considering a polygonal robot $R$ translating and rotating in the plane 
amidst polygonal obstacles. 
A configuration of $R$ describes the position of the reference point 
(center of mass) of $R$  and the orientation of $R$.
As we consider full rotations, the 
configuration space $\calC$ is the three dimensional space $\R^2 \times S^1$. 


\subsecspacea
\subsection{Manifold Families}
\subsecspaceb
As defined in Section \ref{sec:gen_scheme}, we consider manifolds 
defined by \emph{constraints} and construct and decompose them using 
\emph{primitives}. We suggest the following constraints restricting 
motions of $R$ and describe their associated primitives:
The {\bf Angle Constraint} fixes the
orientation of $R$ while it is still free to translate anywhere within the workspace;
the {\bf Segment Constraint} restricts the position of the reference
point to a segment in the workspace while $R$ is free to rotate.

%
%

%
%
%

The left part of Figure \ref{fig:conf_space_slicing} demonstrates 
decomposed manifolds associated to the angle (left bottom) and 
segment (left top) constraints. The angle constraint induces a 
two-dimensional horizontal plane where the cells are polygons. 
The segment constraint induces a two-dimensional vertical slab 
where the cells are defined by the intersection of rational 
curves (as explained in Section \ref{sec:impl_details}). 

We delay the discussion of creating and decomposing manifolds to 
Section~\ref{sec:impl_details}. For now, notice that the 
 Segment-Primitive is far more time-consuming than the Angle-Primitive.

\subsecspacea
\subsection{Exploration and Connection Strategies}
\subsecspaceb
We use manifolds constructed by the Angle-Primitive for the 
exploration phase and manifolds constructed by the Segment-Primitive 
for the connection phase. 
Since the Segment-Primitive is far more costly than the 
Angle-Primitive, we focused our efforts on optimizing the 
former.

\vspace{1mm}
\textbf{Region of interest - RoI:}
As suggested in Section \ref{subsec-roi} we may consider the 
Segment-Primitive in a subset of the range of angles. 
This results in a somewhat ``weaker'' yet more efficient 
primitive than considering the whole range. If the 
connectivity of a local area of the configuration space 
is desired, then using this optimization may suffice 
while considerably speeding up the algorithm.

\vspace{1mm}
\textbf{Generating segments:}
Consecutive layers (manifolds of the Angle Constraint) have
 a similar structure unless topological criticalities occur
 in $\calC$. Once a topological criticality occurs, an 
\FSC either appears and  grows or shrinks  and disappears. 
We thus suggest a heuristic for generating a segment in the
 workspace for the Segment-Primitive using the size of the
 cell as a parameter where we refer to small and large 
cells according to pre-defined constants. The RoI used will 
be proportional to the size of the \FSC. The segment 
generated will be chosen with one of the following 
procedures which are used in Algorithm \ref{alg:choosing_seg}.

\textbf{Random procedure:} Return a random segment from the workspace.

\textbf{Large cell procedure:} Return a random segment in the cell.

\textbf{Small cell procedure:} Intersect the \FSC with the 
next (or the previous) layer. Return a segment connecting a
 random point from the \FSC and a random point in the intersection.

\begin{algorithm}[t,b]
\caption{Generate Segment Constraint ($V$,$E$)}
\label{alg:choosing_seg}
\begin{algorithmic}[1]
	\IF {random\_num ($[0 , 1 ]) \geq $ random\_threshold}
		\RETURN random\_segment\_procedure()
	\ELSE
		\STATE fsc $\leftarrow $ random\_fsc(V)
		\STATE $\alpha \leftarrow $ [size(fsc) - small\_cell\_size] / [large\_cell\_size - small\_cell\_size]
		\IF {random\_num($[0 , 1 ]) \geq \alpha$}
			\RETURN small\_cell\_procedure(fsc,$V$)
		\ELSE
			\RETURN large\_cell\_procedure(fsc,$V$)
		\ENDIF
	\ENDIF
\end{algorithmic}
\end{algorithm}

\vspace{1mm}
\textbf{Constraint Filtering:}
As suggested in Section \ref{subsec-ccf}, we avoid computing 
unnecessary primitives.  All \FSCs that will intersect 
a ``candidate'' constraint $s$, namely all \FSCs of 
layers in its RoI, are tested. If they are all in the 
same connected component in $\calG$, $s$ can be 
discarded as demonstrated in Algorithm \ref{alg:filter_seg}.

\begin{algorithm}[t,b]
\caption{Filter Segment ($s$,$RoI$,$V$,$E$)}
\label{alg:filter_seg}
\begin{algorithmic}[1]
	\STATE $cc_{ids} \leftarrow \emptyset$
	\FORALL{$v \in V$}
		\STATE {fsc $\leftarrow $ free\_space\_cell($v$)}
		\IF {constraining\_angle(fsc)$ \in RoI$}
			\STATE $cc_{ids} \leftarrow cc_{ids} \cup $ connected\_component\_id($v$)
		\ENDIF
	\ENDFOR		
	\IF {$|cc_{ids}| \leq 1$}
		\RETURN filter\_out
	\ENDIF		
\end{algorithmic}
\end{algorithm}

\subsecspacea
\subsection {Path Planning Query}
\subsecspaceb
For a query $q = (q_s,q_t)$, where $q_s = (x_s,y_s,\theta_s)$ 
and $q_t = (x_t,y_t,\theta_t)$, $P_\Theta(\theta_s)$ and 
$P_\Theta(\theta_t)$ are constructed and the \FSCs are 
added to $\calG$. A path of \FSCs between the \FSCs 
containing $q_s$ and $q_t$ is searched for. A local 
path in an Angle-Primitive's \FSC (which is a polygon) 
is constructed by computing the shortest path on the 
visibility graph defined by the vertices of the polygon.
 A local path in an \FSC of a Segment-Primitive 
(which is an arrangement cell) is constructed by 
applying cell decomposition and computing the shortest
 path on the graph induced by the decomposed cells. 
(Figure~\ref{fig:crit} depicts a path in a segment manifold.)
\section{Efficient Implementation of Manifold Decomposition} 
\label{sec:impl_details}
\secspace

The algorithm discussed in Section~\ref{sec:alg_framework} is 
implemented in C++. It is based on \cgal's arrangement package,
 which is used for the geometric primitives, and the \boost 
graph library~\cite{boost-bgl-2001}, which is used to represent 
the connectivity graph \calG.
We next discuss the manifold decomposition methods in more detail. 

\vspace{1mm}
\textbf{Angle-Primitive:}
The Angle-Primitive for a constraining angle $\theta$ 
(denoted $P_\Theta(\theta)$) is constructed by computing 
the Minkowski sum of $-R_\theta$ with the 
obstacles\footnote{$R$ rotated by $\theta$ and reflected about the origin.}. 
The implementation is an application of \cgal's Minkowski sums 
package~\cite[C.24]{cgal}. We remark that we ensure 
(using the method of Canny \etal~\cite{rat-rot-approx}) 
that the angle $\theta$ is chosen such that $\sin\theta$ 
and $\cos\theta$ are rational. This allows for an exact 
rotation of the robot and an exact computation of the Minkowski Sum. 

\vspace{1mm}
\textbf{Segment-Primitive:}
Limiting the possible positions of the robot's reference
point $r$ to a given segment $s$, results in a two-dimensional configuration
space. Each vertex (or edge) of the robot in combination with each edge (or
vertex) of an obstacle gives rise to a critical curve in this configuration space.
Namely the set of all configurations that put the two features into contact, and
thus mark a potential transition between \Cforb and \Cfree. Our analysis (Appendix~\ref{sec:critical_curves}) 
shows that these critical curves can be expressed by rational functions only.
Thus, the implementation of the Segment-Primitive is first of all a computation
of an arrangement of rational functions.

\cgal follows the \emph{generic programming paradigm}~\cite{a-gps-98}, 
that is, algorithms are formulated and implemented such that they 
abstract away from the actual types, constructions and predicates.
Using the \CC programming language this is realized
by means of class and function templates, respectively.
In particular, the arrangement package is written such that it takes
a traits class as a template argument. This traits class defines the supported
curve type and provides the operations that are required for this type. 
Since the old specialized traits class was too slow 
(even slower than the solution for general algebraic curves presented in~\cite{bhk-gak-2011}), 
we devised a new efficient
traits class for rational functions.

The new traits class (all details in Appendix~\ref{sec:traits}) is written such that it takes
maximal advantage of the fact that the supported curves are 
functions. As opposed to the general traits in~\cite{bhk-gak-2011}, 
we never have to shear the coordinate system
and we only require tools provided by the univariate algebraic kernel of \cgal~\cite[C.8]{cgal}.
A comparison using the benchmark instances that were also used in~\cite{bhk-gak-2011} shows
that the new traits class is about 3-4 times faster then the general traits class;
this is a total speed up of about 10 when compared to the old dedicated traits class.

The development of this new traits class represents the low
tier of our efforts to produce an effective motion planner
and relies on a more intimate acquaintance with \cgal in
general and arrangement traits for algebraic curves in
particular; therefore we defer further details to the appendix.
We note that the new traits class has been accepted for 
integration into \cgal and will be available in the upcoming 
\cgal release~3.9.  

\section{Experimental Results }
\label{sec:exp_results}
\begin{figure*}[t]
  \centering
  \subfloat
  	[Tunnel scenario \vspace{-3mm}]
  	{\label{fig:scenario_1}
  	\includegraphics[width=0.28\textwidth]{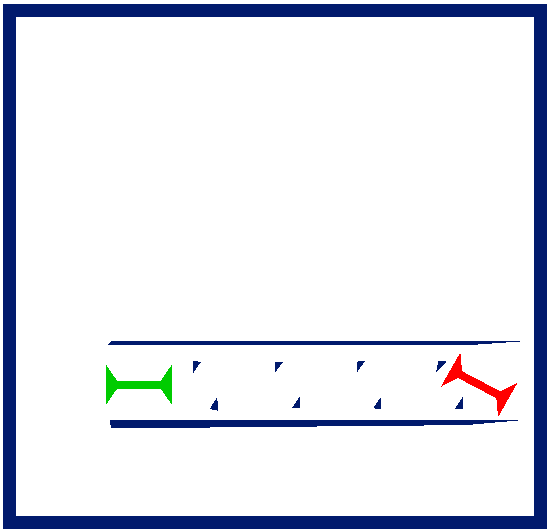}}
  \subfloat
  	[Snake scenario \vspace{-3mm}]
  	{\label{fig:scenario_2}
  	\includegraphics[width=0.28\textwidth]{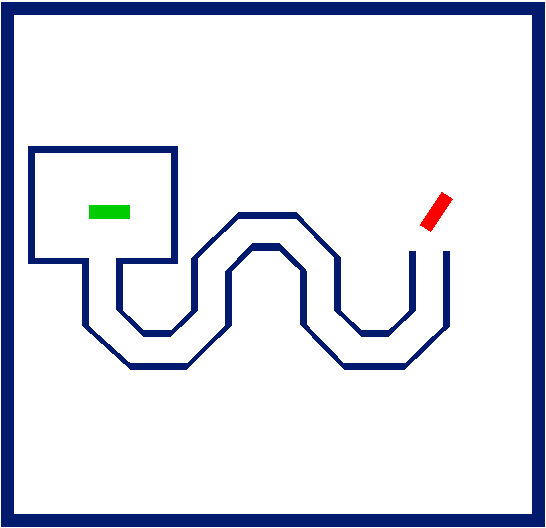}}
  \subfloat						
  	[Flower scenario  \vspace{-3mm}]
  	{\label{fig:scenario_3}
  	\includegraphics[width=0.28\textwidth]{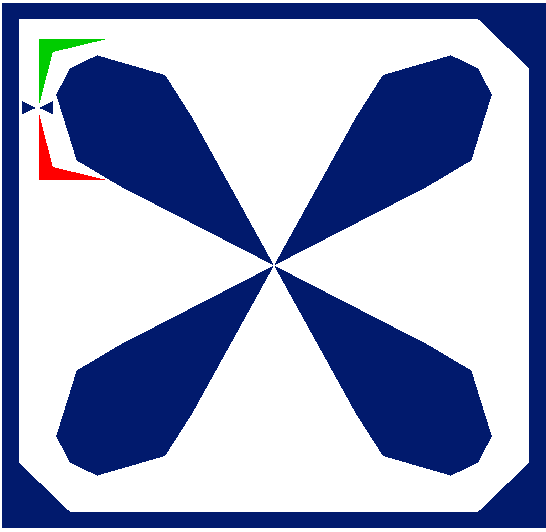}}
  \caption{Experimental scenarios}
  \label{fig:scenarios}
 \vspace{-6mm}
\end{figure*}

\secspace
We demonstrate the performance of our planner using three different scenarios. All scenarios consist of a workspace, a robot with obstacles and one query (source and target configurations). Figure \ref{fig:scenarios} illustrates the scenarios where the obstacles are drawn in blue and the source and target configurations are drawn in green and red, respectively. All reported tests were measured on a Dell 1440 with one 2.4GHz P8600 Intel Core 2 Duo CPU processor and 3GB of memory  running with a Windows 7 32-bit OS. Preprocessing times presented are times that yielded at least 80\% (minimum of 5 runs) success rate in solving queries.

\subsecspacea
\subsection{Algorithm Properties}
\subsecspaceb
Our planner has two parameters: the number $n_\theta$ of layers to be generated and the number $n_s$ of segment constraints to be generated. We chose the following values for these parameters: $n_\theta \in \{10,20,40,80\}$ and $n_s \in \{2^i | i \in \N, i \leq 14\}$. 
For a set of parameters $(n_\theta,n_s)$ we report the preprocessing time $t$ and whether
a path was found (marked \niceV) or not found (marked \niceX) once the query was issued.
The results for the flower scenario are reported in Table \ref{tbl:params-sensistivity}. We show that a considerable increase in parameters has only a limited effect on the preprocessing time. 

In order to test the effectiveness of our optimizations, we ran the planner with and without any heuristic for choosing segments and with and without segment filtering.  We also added a test with all optimizations using the old traits class. The results for the flower scenario can be viewed in Table \ref{tbl:optimization}. We remark that the engineering work invested in optimizing \MMS yielded an algorithm comparable and even surpassing a motion planner that is in prevalent use as shown next.

\begin{table}[t]
\parbox{.4\linewidth}
{
	\begin{center}
	\vspace{-5mm}
	\begin{tabular}{cc|c|c|c|c|}
		\cline{3-6}
		& & \multicolumn{4}{|c|}{$n_\theta$} \\ \cline{3-6}
		& 												&		10	 			& 20					& 40					& 80 					\\ \cline{1-6}
		\multicolumn{1}{|c|}{\multirow{4}{*}{$n_s$}} &
		\multicolumn{1}{|c|}{256}	& (6,\niceX)	& (11,\niceX)	& (12,\niceX)	& (16,\niceX)  \\ \cline{2-6}
		\multicolumn{1}{|c|}{}                        &
		\multicolumn{1}{|c|}{512} & (7,\niceX)  & (13,\niceX)	& (14,\niceX)	& (25,\niceX)  \\ \cline{2-6}
		\multicolumn{1}{|c|}{} &
		\multicolumn{1}{|c|}{1024}& (16,\niceX)	& (20,\niceV) & (23,\niceV) & (35,\niceV)  \\ \cline{2-6}
		\multicolumn{1}{|c|}{}                        &	
		\multicolumn{1}{|c|}{2048}& (30,\niceX)	& (35,\niceV) & (38,\niceV)	& (51,\niceV)		\\ \cline{1-6}
		\multicolumn{1}{|c|}{}                        &	
		\multicolumn{1}{|c|}{4096}& (46,\niceV)	& (53,\niceV) & (60,\niceV)	& (82,\niceV)		\\ \cline{1-6}
	\end{tabular}
	\end{center}
	\caption{\label{tbl:params-sensistivity} Parameter sensitivity}
	\vspace{-10mm}
}
\hfill
\parbox{.5\linewidth}
{
	\begin{center}
	\vspace{-5mm}
	\begin{tabular}{|c|c|c|c|c||c|}
		\hline
													&Segment 											&						&							&						&	 		\\
		Traits								&	Generation 									& Filtering &	$n_\theta$	&	$n_s$	    &	 t	\\ 	\hline
		\multirow{4}{*}{New}	&	\multirow{2}{*}{random}			& not used	&			20			&	8192			&	1418\\	\cline{3-6}
	 												&															& used			&			20			&	8192			&	112	\\	\cline{2-6}
													&\multirow{2}{*}{heuristic}		& not used	&			40			&	512				&	103	\\	\cline{3-6}
 													&															& used			&			20			&	1024			&	20	\\	\hline 		
 		Old										&heuristic										& used			&			20			&	1024			&	138	\\	\hline 		
		\end{tabular}
	\end{center}
	\caption{\label{tbl:optimization} Optimization results}
	\vspace{-10mm}
}
\end{table}

\subsecspacea
\subsection{Comparison With PRM}
\begin{figure}[b]
\parbox{.5\linewidth}
{
	\captionsetup{type=table}
	\begin{center}
	\vspace{-5mm}
	\vspace{3mm}
	\begin{tabular}{|c||c|c|c||c|c|c||c|}
	\hline
	Scenario 	& \multicolumn{3}{|c||}{MMS} 							& \multicolumn{3}{|c||}{PRM} 									& Speedup \\\cline{2-7}
						&  $n_\theta$	& $n_s$       & t 					& k						& \% st						& t						&					\\\hline
	Tunnel	 	&		20				&	512					&	100					&	20					&	0.0125					&	180					&		1.8		\\
	Snake			&		40				&	256					&	22					&	20					&	0.025						&	140					&		6.3		\\
	Flower		&		20				&	1024				&	20					&	24					&	0.0125					&	40					&		2			\\\hline
	\end{tabular}
	\end{center}
  \vspace{3mm}
	\caption{\label{tbl:results1} Comparison With PRM  }
	\vspace{-4mm}
}
\hfill
\parbox{.45\linewidth}
{
	  \centering
	  \vspace{-5mm}
   	\includegraphics[width=0.35\textwidth]{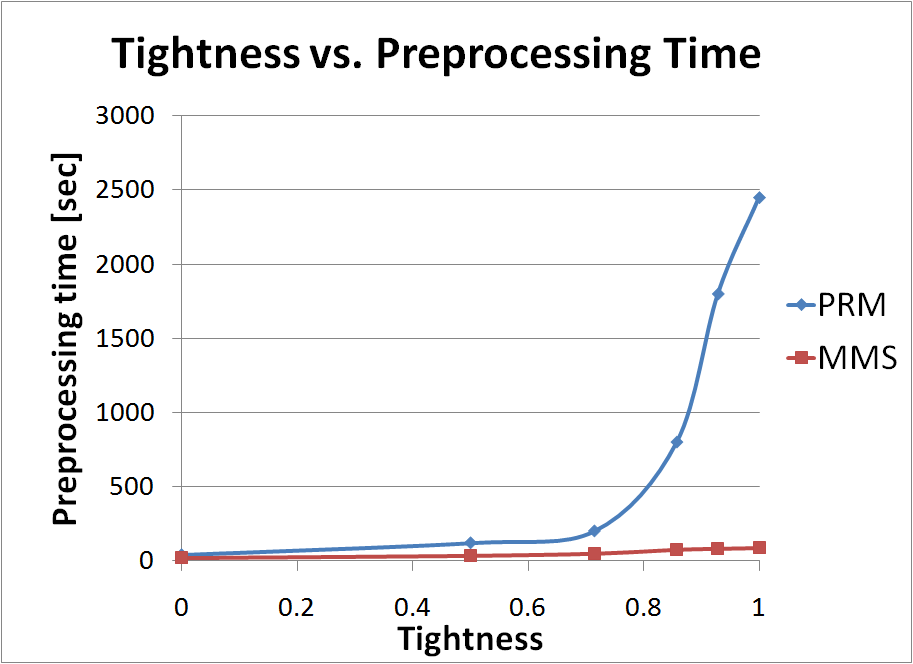}
   	\caption{\label{fig:tightness}Tightness Results}
   	\vspace{-5mm}
}
\end{figure}

\subsecspaceb
We used an implementation of the PRM algorithm as provided by the OOPSMP package~\cite{OOPSMP}. For fair comparison, we did not use cycles in the roadmap as cycles increase the preprocessing
 time significantly. We manually optimized the parameters of each planner over a concrete set. As with previous tests, the parameters for \MMS are $n_\theta$ and $n_s$. The parameters used for the PRM are the number of neighbors (denoted $k$) to which each milestone should be connected and the percentage of time used to sample new milestones (denoted \% st in Table \ref{tbl:results1}).

Furthermore, we ran the flower scenario several times, progressively increasing the robot size. This caused a ``tightening'' of the passages containing the desired path. Figure \ref{fig:tightness} demonstrates the preprocessing time as a function of the tightness of the problem for both planners. A tightness of zero denotes the base scenario (Figure \ref{fig:scenario_3}) while a tightness of one denotes the tightest problem solved. 

The results show a speedup for all scenarios when compared to the PRM implementation. Moreover, our algorithm has little sensitivity to the tightness of the problem as opposed to the PRM algorithm. In the tightest experiment, \MMS runs 27 times faster than the PRM implementation.

\section{Further Directions}
\label{sec:conclusion}
\secspace
To conclude, we outline directions for extending
and enhancing the current work. Our primary goal is to use
the \MMS framework to solve progressively more complicated
motion-planning problems. As suggested earlier, we see the
framework as a platform for convenient transfer of strong
geometric primitives into motion planning algorithms. For
example, among the recently developed tools are efficient
and exact solutions for computing the Minkowski sums of
polytopes in $\R^3$ (see Introduction) as well as for
exact update of the sum when the polytopes rotate~\cite{MFH10}. 
These could be combined
into an \MMS for planning full rigid motion of a polytope
among polytopes, which, extrapolating from the current
experiments could outperform more simplistic solutions in
existence.

Looking at more intricate problems, we anticipate some
difficulty in turning constraints into manifolds that can
be exactly decomposed. We propose to have manifolds where
the decomposition yields some {\it approximation} of the \FSCs,
using recent advanced meshing tools for example. We can
endow the connectivity-graph nodes with an attribute
describing their approximation quality. One can then
decide to only look for paths all whose nodes are above
a certain approximation quality. Alternatively, one can
extract any solution path and then refine only those
portions of the path that are below a certain quality.


\textbf{Beyond motion planning:}  We foresee an extension of the framework to other problems that involve high-dimensional arrangements of critical hypersurfaces. It is difficult to describe the entire arrangement analytically, but there are often situations where constraint manifolds could be computed analytically. Hence, it is possible to shed light on problems such as loop closure and assembly planning where we can use manifold samples to analytically capture pertinent information of high-dimensional arrangements of hypersurfaces. Notice that although in Section~\ref{sec:alg_framework} we used only planar manifolds, there are recently developed tools to construct two dimensional arrangement of curves on curved surfaces~\cite{BerberichFHMW10} which gives further flexibility in choosing the manifold families.

For supplementary material and updates the reader is referred to our webpage \url{http://acg.cs.tau.ac.il/projects/mms}.


\bibliography{bibliography}

\begin{thebibliography}{10}

\bibitem{AronovS97}
B.~Aronov and M.~Sharir.
\newblock On translational motion planning of a convex polyhedron in 3-space.
\newblock {\em SIAM J. Comput.}, 26(6):1785--1803, 1997.

\bibitem{a-gps-98}
M.~H. Austern.
\newblock {\em Generic Programming and the {STL}}.
\newblock Addison-Wesley, 1998.

\bibitem{AvnaimBF88}
F.~Avnaim, J.-D. Boissonnat, and B.~Faverjon.
\newblock A practical exact motion planning algorithm for polygonal object
  amidst polygonal obstacles.
\newblock In {\em Proceedings of the Workshop on Geometry and Robotics}, pages
  67--86, Springer-Verlag, 1989.

\bibitem{Basu03}
S.~Basu, R.~Pollack, and M.-F. Roy.
\newblock {\em Algorithms in Real Algebraic Geometry}.
\newblock Algorithms and Computation in Mathematics. Springer-Verlag, 2003.

\bibitem{BerberichFHMW10}
E.~Berberich, E.~Fogel, D.~Halperin, K.~Mehlhorn, and R.~Wein.
\newblock Arrangements on parametric surfaces \uppercase{I}: General framework
  and infrastructure.
\newblock {\em Mathematics in Computer Science}, 4(1):45--66, 2010.

\bibitem{bhk-gak-2011}
E.~Berberich, M.~Hemmer, and M.~Kerber.
\newblock A generic algebraic kernel for non-linear geometric applications.
\newblock In {\em SoCG 2011}.

\bibitem{rat-rot-approx}
J.~Canny, B.~Donald, and E.~K. Ressler.
\newblock A rational rotation method for robust geometric algorithms.
\newblock In {\em SoCG 1992}, pages 251--260, ACM.

\bibitem{Canny-complexity}
J.~F. Canny.
\newblock {\em Complexity of Robot Motion Planning (ACM Doctoral Dissertation
  Award)}.
\newblock The MIT Press, June 1988.

\bibitem{Chazelle199177}
B.~Chazelle, H.~Edelsbrunner, L.~J. Guibas, and M.~Sharir.
\newblock A singly exponential stratification scheme for real semi-algebraic
  varieties and its applications.
\newblock {\em Theoretical Computer Science}, 84(1):77 -- 105, 1991.

\bibitem{planning-survey-choset}
H.~Choset, W.~Burgard, S.~Hutchinson, G.~Kantor, L.~E. Kavraki, K.~Lynch, and
  S.~Thrun.
\newblock {\em Principles of Robot Motion: Theory, Algorithms, and
  Implementation}.
\newblock MIT Press, June 2005.

\bibitem{CG-alg-app}
M.~De~Berg, O.~Cheong, M.~van Kreveld, and M.~Overmars.
\newblock {\em {Computational Geometry: Algorithms and Applications}}.
\newblock Springer, 2008.

\bibitem{fogel-mink}
E.~Fogel and D.~Halperin.
\newblock Exact and efficient construction of \uppercase{M}inkowski sums of
  convex polyhedra with applications.
\newblock {\em CAD}, 39(11):929--940, 2007.

\bibitem{Hachenberger09}
P.~Hachenberger.
\newblock Exact \uppercase{M}inkowksi sums of polyhedra and exact and efficient
  decomposition of polyhedra into convex pieces.
\newblock {\em Algorithmica}, 55(2):329--345, 2009.

\bibitem{HS96}
D.~Halperin and M.~Sharir.
\newblock A near-quadratic algorithm for planning the motion of a polygon in a
  polygonal environment.
\newblock {\em Disc. Comput. Geom.}, 16(2):121--134, 1996.

\bibitem{hybrid-disks}
S.~Hirsch and D.~Halperin.
\newblock Hybrid motion planning: Coordinating two discs moving among polygonal
  obstacles in the plane.
\newblock In {\em WAFR 2002}, pages 225--241.

\bibitem{Hsu99}
D.~Hsu, J.~Latombe, and R.~Motwani.
\newblock Path planning in expansive configuration spaces.
\newblock {\em Int. J. Comp. Geo. \& App.}, 4:495--512, 1999.

\bibitem{Kavraki98}
L.~E. Kavraki, M.~N. Kolountzakis, and J.-C. Latombe.
\newblock Analysis of probabilistic roadmaps for path planning.
\newblock {\em IEEE Trans. Robot. Automat.}, 14(1):166--171, 1998.

\bibitem{Kavraki-PRM}
L.~E. Kavraki, P.~Svestka, J.-C. Latombe, and M.~Overmars.
\newblock Probabilistic roadmaps for path planning in high dimensional
  configuration spaces.
\newblock {\em IEEE Transactions on Robotics and Automation}, 12(4):566--580,
  1996.

\bibitem{Kuffner00}
J.~J. Kuffner and S.~M. Lavalle.
\newblock {RRT}-{C}onnect: An efficient approach to single-query path planning.
\newblock In {\em ICRA 00'}, pages 995--1001, 2000.

\bibitem{LK02}
A.~M. Ladd and L.~E. Kavraki.
\newblock Generalizing the analysis of {PRM}.
\newblock In {\em Proceedings of the 2002 IEEE International Conference on
  Robotics and Automation (ICRA 2002)}, pages 2120--2125, IEEE Press, 2002.

\bibitem{Latombe-Robot-Mot-Plan}
J.-C. Latombe.
\newblock {\em Robot Motion Planning}.
\newblock Kluwer Academic Publishers, Norwell, MA, USA, 1991.

\bibitem{Lavalle-RRT}
S.~M. Lavalle.
\newblock Rapidly-exploring random trees: A new tool for path planning.
\newblock {\em In Computer Science Dept., Iowa State University}, Tech.
  Rep:98--11, 1998.

\bibitem{planning-survey-lavalle}
S.~M. LaValle.
\newblock {\em Planning Algorithms}.
\newblock Cambridge University Press, Cambridge, U.K., 2006.

\bibitem{hybrid-mink}
J.-M. Lien.
\newblock Hybrid motion planning using {M}inkowski sums.
\newblock In {\em RSS 2008}.

\bibitem{Lozano-Perez80spatialplanning}
T.~Lozano-Perez.
\newblock Spatial planning: A configuration space approach.
\newblock {\em {MIT} {AI} {M}emo 605}, 1980.

\bibitem{MFH10}
N.~Mayer, E.~Fogel, and D.~Halperin.
\newblock Fast and robust retrieval of {M}inkowski sums of rotating convex
  polyhedra in 3-space.
\newblock In {\em SPM}, pages 1--10, 2010.

\bibitem{OOPSMP}
E.~Plaku, K.~E. Bekris, and L.~E. Kavraki.
\newblock {OOPS} for motion planning: An online open-source programming system.
\newblock In {\em ICRA}, pages 3711--3716, IEEE, 2007.

\bibitem{Piano-complexity}
J.~H. Reif.
\newblock Complexity of the mover's problem and generalizations.
\newblock In {\em FOCS}, pages 421--427, IEEE Computer Society, 1979.

\bibitem{PianoI}
J.~T. Schwartz and M.~Sharir.
\newblock On the "piano movers" problem: I. \uppercase{T}he case of a
  two-dimensional rigid polygonal body moving amidst polygonal barriers.
\newblock {\em Commun. Pure appl. Math}, 35:345 -- 398, 1983.

\bibitem{PianoII}
J.~T. Schwartz and M.~Sharir.
\newblock On the "piano movers" problem: \uppercase{ii}. \uppercase{G}eneral
  techniques for computing topological properties of real algebraic manifolds.
\newblock {\em Advances in Applied Mathematics}, 4(3):298 -- 351, 1983.

\bibitem{planning-survey-sharir}
M.~Sharir.
\newblock {\em Algorithmic Motion Planning, Handbook of Discrete and
  Computational Geometry}.
\newblock 2nd Edition, CRC Press, Inc., Boca Raton, FL, USA, 2004.

\bibitem{boost-bgl-2001}
J.~G. Siek, L.-Q. Lee, and A.~Lumsdaine.
\newblock {\em The Boost Graph Library: User Guide and Reference Manual}.
\newblock Addison-Wesley Professional, 2001.

\bibitem{cgal}
{The CGAL Project}.
\newblock {\em {CGAL} User and Reference Manual}.
\newblock {CGAL Editorial Board}, {3.7} edition, 2010.
\newblock http\://www.cgal.org/.

\bibitem{w-eecpm-06}
R.~Wein.
\newblock Exact and efficient construction of planar {M}inkowski sums using the
  convolution method.
\newblock In {\em ESA}, pages 829--840, 2006.

\bibitem{YangS06}
J.~Yang and E.~Sacks.
\newblock \uppercase{RRT} path planner with 3 {DOF} local planner.
\newblock In {\em ICRA}, pages 145--149, 2006.

\bibitem{ZhangKM07}
L.~Zhang, Y.~J. Kim, and D.~Manocha.
\newblock A hybrid approach for complete motion planning.
\newblock In {\em IROS}, pages 7--14, 2007.

\end{thebibliography}
\bibliographystyle{abbrv-short}

\pagebreak
\appendix
\section{Traits Class for Rational Functions}
\label {sec:traits}

For completeness we repeat below (in Section~\ref{subsec:traits-bkgnd}) material 
that already appears in the body of the paper. More technical details on the 
implementation of the traits class are given in Section~\ref{subsec:traits-impl}.

\subsection{Background}
\label{subsec:traits-bkgnd}
\cgal follows the \emph{generic programming paradigm}~\cite{a-gps-98}, 
that is, algorithms are formulated and implemented such that they 
abstract away from the actual types, constructions, and predicates.
Using the \CC programming language this is realized
by means of class and function templates, respectively.
The manifold decomposition methods described
below use \cgal's arrangement package.

As most other high level packages of \cgal, the arrangement 
package is written such that it takes one \emph{traits} class 
as a template argument. This traits class defines the type of 
curves in use and the required operations on them.
This gives tremendous flexibility  in using the package for a
variety of different motion planning (sub)problems. Here we
use it once for line segments via the Minkowski sums package
written on top of the arrangement package, and once with
graphs of rational functions, for which we devised a new and
particularly efficient traits class. 
The following section gives more detail about the later. 

The new traits class is written such that it takes maximal
advantage of the fact that the supported curves are
functions. This implies that, as opposed to
the general traits in~\cite{bhk-gak-2011}, we never have to shear the
coordinate system and that we only require tools provided
by the univariate algebraic kernel~\cite[C.8]{cgal}. 

A traits class is required to provide (and by that define) the 
curve type in use. For these curves it must provide a specific set 
of operations, such as splitting curves into $x$-monotone sub-curves,
computing the intersections of two curve segments, 
comparing intersection points,
or comparing the $y$-order of two curves at a certain $x$-coordinate.  
Thus, our traits class implements these predefined operations explicitly 
for rational functions. 
The heart of the traits consists of two classes that 
represent the complete topology of one or two functions, respectively. 
All predicates and constructions required for the traits are
essentially just queries to one of these two classes. 

\subsection{Technical Details}
\label{subsec:traits-impl}
For one function $f(x) = f_n(x)/f_d(x)$, with $f_n, f_d\in\Z[X]$ coprime, 
we subdivide the $x$-axis into intervals such that the sign 
of $f$ is invariant within each interval. 
This is obtained by computing the sorted sequence of the real roots 
(with multiplicity) of $f_n$ and $f_d$.
For the right-most interval the sign is determined by the signs of the 
leading coefficients of $f_n$ and $f_d$. The remaining signs 
are concluded from right to left using the multiplicity 
of the computed roots. 

For two functions $f(x)= f_n(x)/f_d(x)$ and $g(x)= g_n(x)/g_d(x)$,
we similarly subdivide the $x$-axis into intervals such that the  
$y$-order of $f$ and $g$ is invariant within each interval. 
Of course, the order may change at intersection points, the  
$x$-coordinates of which are given by the real roots of $r=f_ng_d - g_nf_d \in \Z[X]$.
However, the order may also change at vertical asymptotes of $f$ and $g$. 
Thus, the subdivision is given by the sorted sequence of the real roots
of $f_d$, $g_d$ and $r$. 
Again, once the order is computed for one interval, the 
others can be concluded via the multiplicities of the roots. 
All results are cached such that each instance of 
both classes is computed at most once. 

As stated in Section~\ref{sec:impl_details} comparison with the benchmark instances that were also used 
in~\cite{bhk-gak-2011} shows that the new traits class is about 
3-4 times faster then the general traits class, this is a 
total speed up of about 10 when compared to the old dedicated 
traits class, which was based on \core. 
We remark that the new traits class has already been accepted for 
integration into \cgal and will be available in the upcoming 
\cgal release~3.9.  

\section{Critical Curves of a Rotating Robot Along a Translation Segment}
\label{sec:critical_curves}

\subsection{The problem}
\begin{figure}
  \vspace{-3mm}
  \centering
   \begin{center}
    \includegraphics[width=0.35\textwidth]{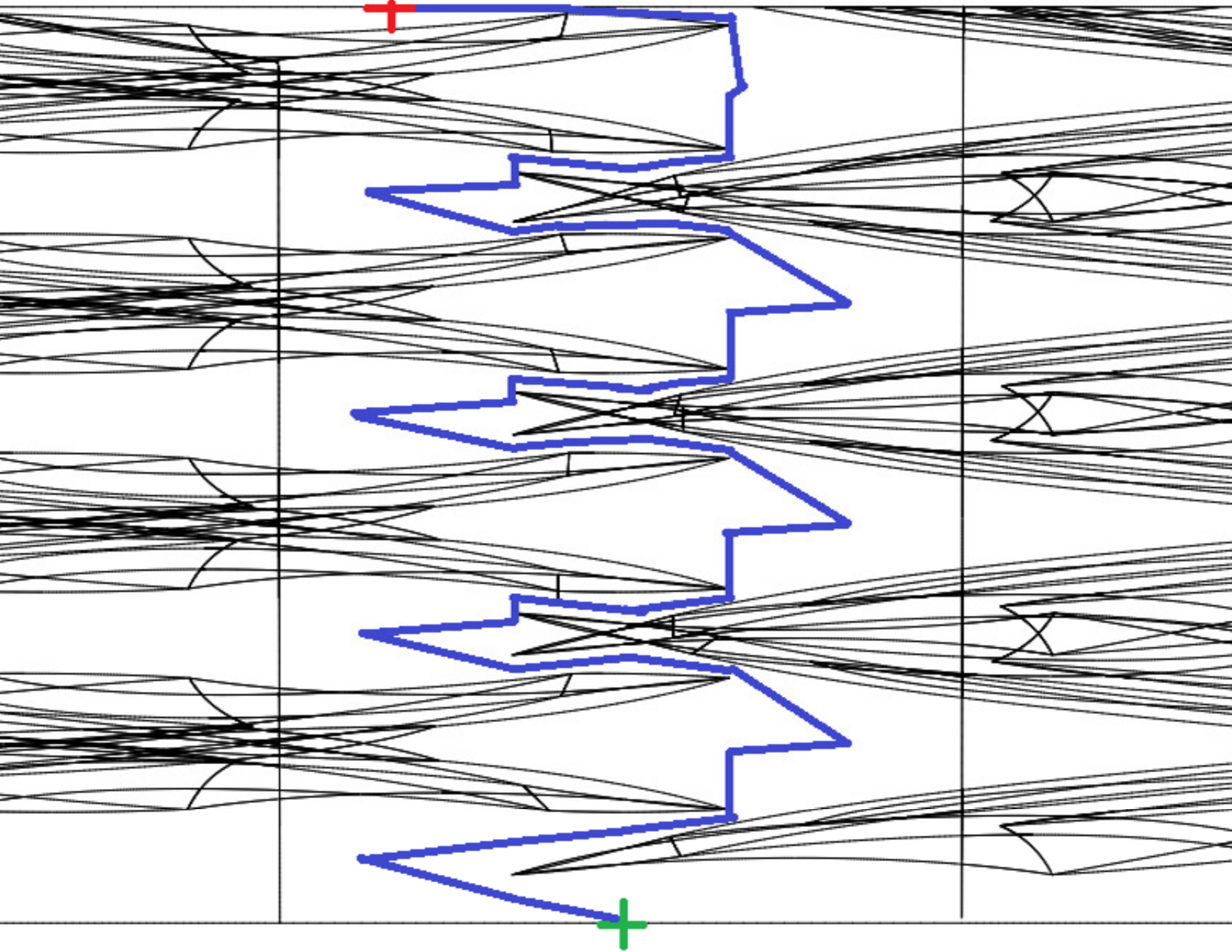}
   \end{center}
   \vspace{-2mm}
   \caption{Arrangement of critical curves}
   \label{fig:crit}
   \vspace{-2mm}
\end{figure}

We consider a polygonal robot translating along a fixed segment while rotating amidst polygonal obstacles. This means that the reference point of the robot moves along a fixed line segment, and the robot can rotate around this reference point. A critical curve is a curve representing a motion of the robot while a feature of the robot is in contact with a feature of an obstacle: Either a robot's vertex is in contact with an obstacle's edge or a robot's edge is in contact with an obstacle's vertex. These cases will be referred to as \emph{vertex-edge} and \emph{edge-vertex} respectively. Figure~\ref{fig:crit} demonstrates part of an arrangement of critical curves constructed for the Tunnel scenario (Figure~\ref{fig:scenario_1}) for a segment connecting the source and target configurations. The green and red crosses mark the source and target configurations respectively and the blue poly-line is a path constructed within an \FSC.

We define the critical curves by first assuming that both the segment and the edge at hand (either the robot's or the obstacle's) is a full line (containing the edge). We then introduce ``critical endpoints'' taking into account the fact that the robot's translation is restricted to a segment and that the edge considered is not a line. This is done by identifying the geometric locus where a vertex and an edge's endpoint are in contact. We omit this additional stage from the appendix though we addressed it in our work. 
 
\subsection{Robot representation and definitions}
A robot $R$ is a simple polygon with vertices $\{v_1,\dots,v_n\}$ 
where  $v_i = (x_i,y_i)^T$
and edges $\{(v_1,v_2),  \dots (v_n,v_1)\}$.
We assume that the reference point of $R$ is located at the origin. 
The position of $R$ in the workspace is defined by a configuration $q = (r_q,\theta_q) $  
where $r_q = (x_q,y_q)^T$.  

Thus, $q$ maps the position of a vertex $v_i$ by the following parameterization: 
\[ v_i(q) = M(\theta_q)v_i+r_q , \]
where 
$M(\theta) = 
 \left[ 
  \begin{array}{cc} 
   \cos\theta & -\sin\theta \\ 
   \sin\theta & \cos\theta
  \end{array} 
 \right]$
is the rotation matrix. 

Given a segment $seg = [s,t]$ where 
$s = (x_s,y_s)^T$ 
and
$t = (x_t,y_t)^T$
we define the parametrization $(\alpha,\tau ) \in [0,1] \times \R$ in Equations (\ref{eq:par_r}), (\ref{eq:par_T})
\begin{equation}
 r_q = (1-\alpha)s+\alpha t,
 \label{eq:par_r}
\end{equation}

\begin{equation}
 \theta_q = 2 \arctan \tau .
 \label{eq:par_T}
\end{equation}

The parametrization fixes the robot's reference point to the supporting line of $seg$. The parametrized vertex is represented in Equation (\ref{eq:v(q)}) 
\begin{equation}
 v_i(\alpha,\tau ) = M(\tau )v_i + (1-\alpha)s+\alpha t,
 \label{eq:v(q)}
\end{equation}

where
$M(\tau ) = \frac{1}{1+\tau^2} 
 \left[ 
  \begin{array}{cc} 
   1-\tau^2 & -2\tau   \\ 
   2\tau   & 1-\tau^2
  \end{array} 
 \right]$.

\subsection{Robot's vertex - Obstacle's edge}
Let $v_i $ be a robot's vertex and $e$ be an obstacle's edge that is supported by the line $l: ax + by + c = 0$. The critical curve in this case is defined by adding the constraint that $v_i(q) \in l$ thus:
\begin{equation}
 ax_i(q) + by_i(q) + c = 0.
 \label{eq:obs_line}
\end{equation}

Plugging Equation (\ref{eq:v(q)}) into Equation (\ref{eq:obs_line}) yields: \\
$a [(1-\tau^2)x_i-2\tau y_i] + a(1+\tau^2)(1-\alpha)x_s + a(1+\tau^2) \alpha x_t + $ \\
$b [(2\tau )x_i+(1-\tau^2)y_i] + b(1+\tau^2)(1-\alpha)y_s + b(1+\tau^2)\alpha y_t + $ \\
$c(1+\tau^2) = 0$, \\
when simplified: \\ 
$\tau^2     \cdot [- a x_i + a x_s - b y_i + b y_s + c] + $\\
$\tau       \cdot [ - 2 a y_i + 2b x_i ] + $
$1      \cdot [a x_i + a x_s + b y_i + b y_s + c] + $  \\
$\alpha \tau^2 \cdot [-a x_s + a x_t - b y_s + b y_t ] + $
$\alpha   \cdot [- a x_s + a x_t - b y_s + b y_t ] = 0 $.

Hence:
\begin{equation}
 \alpha = \frac{p_2 \tau^2 + p_1 \tau  + p_0}{q_2 \tau^2 + q_0} ,
\end{equation}

where \\ \\
\begin{minipage}{0.4\linewidth}
 $p_2 = a (x_i - x_s) + b ( y_i - y_s) - c$,\\
 $p_1 = 2 ( a y_i - b x_i )$,\\
 $p_0 = -a ( x_i + x_s ) - b ( y_i + y_s ) - c$,\\
\end{minipage}
\hfill
\begin{minipage}{0.4\linewidth}
$q_2 = a ( x_t - x_s ) + b ( y_t - y_s)$,\\
\\
$q_0 = q_2$.\\
\end{minipage}
 
\subsection{Robot's edge - Obstacle's vertex}
Let $e = (v_1,v_2) $ be a robot's edge and $v_0$ be an obstacle's vertex. The critical curve in this case is defined by adding the constraint that $v_0$ lies on the line $l : ax + by +c = 0$ supporting $e$. To simplify the notation we will consider $\widetilde{l} : (1+\tau^2)ax + (1+\tau^2)by + (1+\tau^2)c = 0$ and the constraint that $v_0 \in \widetilde{l}$. This line has the following coefficients:
$(1+\tau^2)a = (1+\tau^2)(y_2(q) - y_1(q))$, $(1+\tau^2)b = (1+\tau^2)(x_1(q) - x_2(q))$ and $(1+\tau^2)c = (1+\tau^2)(x_2(q) y_1(q) - x_1(q) y_2(q))$.
Let us denote $\Delta_x = (x_2 - x_1)$ and $\Delta_y = (y_2 - y_1)$. \\

Simplifying the coefficients of $\widetilde{l}$ yields: \\ 

\begin{table}[H]
\begin{tabular}{lcl}
$(1+\tau^2)a$ &  =  &  $(1+\tau^2)(y_2(q) - y_1(q))$ \\
      &  =  &  $[(2\tau )x_2+(1-\tau^2)y_2] + (1+\tau^2)[(1-\alpha)y_s + \alpha y_t]  $ \\
      &    &  $-[(2\tau )x_1+(1-\tau^2)y_1] - (1+\tau^2)[(1-\alpha)y_s + \alpha y_t]  $ \\
      &  =  &  $[\Delta_y + (2\Delta_x)\tau  - \Delta_y \tau^2]      $,\\
\\
$(1+\tau^2)b$ &  =  &  $(1+\tau^2)(x_1(q) - x_2(q))$ \\
      &  =  &  $[(1-\tau^2)x_1-2\tau y_1] + (1+\tau^2)[(1-\alpha)x_s + \alpha x_t]  $ \\
      &    &  $-[(1-\tau^2)x_2-2\tau y_2] - (1+\tau^2)[(1-\alpha)x_s + \alpha x_t]  $ \\
      &  =  &  $[-\Delta_x + \Delta_y 2\tau  + \Delta_x \tau^2]     $,\\
\\
$(1+\tau^2)c$ &  =  &  $(1+\tau^2)(x_2(q) y_1(q) - x_1(q) y_2(q))$\\

      &  =  &  $  (1+\tau^2)[\frac{1}{1+\tau^2} [(1-\tau^2)x_2-2\tau y_2]   + (1-\alpha)x_s + \alpha x_t]$ \\
      &    &  $      [\frac{1}{1+\tau^2} [(2\tau )x_1+(1-\tau^2)y_1] + (1-\alpha)y_s + \alpha y_t]$ \\
      &    &  $- (1+\tau^2)[\frac{1}{1+\tau^2} [(1-\tau^2)x_1-2\tau y_1]   + (1-\alpha)x_s + \alpha x_t]$\\
      &    &  $     [\frac{1}{1+\tau^2} [(2\tau )x_2+(1-\tau^2)y_2] + (1-\alpha)y_s + \alpha y_t]$\\

      &  =  &  $\frac{1}{(1+\tau^2)} [-4 (x_1 y_2  - x_2 y_1) \tau^2 - (x_1 y_2 - x_2 y_1) (1-\tau^2)^2 ]  $\\
      &    &  $+          [(1-\tau^2)\Delta_x - 2\tau \Delta_y] [y_s + (y_t - y_s) \alpha]     $\\
      &    &  $+          [-2\tau \Delta_x - (1-\tau^2)\Delta_y] [x_s + (x_t - x_s) \alpha]   $\\

\end{tabular}
\end{table}

\begin{table}[H]
\begin{tabular}{lcl}
      
$(1+\tau^2)c$ &  =  &  $-(1+\tau^2)(x_1 y_2 - x_2 y_1)                     $\\
      &    &  $+ ( \Delta_x - 2\Delta_y \tau  - \Delta_x \tau^2) (y_s + (y_t - y_s) \alpha)$\\
      &    &  $+ (-\Delta_y - 2\Delta_x \tau  + \Delta_y \tau^2) (x_s + (x_t - x_s) \alpha)$.\\
      
\end{tabular}
\end{table}

Denoting $k = x_1 y_2 - x_2 y_1 $ and inserting $v_0$ into the line equation of $\widetilde{l}$yields:\\
$ [\Delta_y + (2\Delta_x)\tau  - \Delta_y \tau^2] x_0 $ 
$+[-\Delta_x + \Delta_y 2\tau  + \Delta_x \tau^2] y_0 $ 
$- k (1+\tau^2)                  $ \\
$+[ \Delta_x - 2\Delta_y \tau  - \Delta_x \tau^2] [y_s + (y_t - y_s) \alpha]    $ 
$+[-\Delta_y - 2\Delta_x \tau  + \Delta_y \tau^2] [x_s + (x_t - x_s) \alpha]  = 0 $. \\ 

Now,\\
$\tau^2     \cdot [  \Delta_x (y_0 - y_s) -   \Delta_y (x_0 - x_s) - k] +  $\\ 
$\tau       \cdot [2  \Delta_x (x_0 - x_s) + 2 \Delta_y (y_0 - y_s)] +    $
$1      \cdot [-  \Delta_x (y_0 - y_s) +   \Delta_y (x_0 - x_s) - k] +  $\\ 
$\alpha \tau^2 \cdot [- \Delta_x (y_t - y_s) +   \Delta_y (x_t - x_s)] +    $\\
$\alpha \tau    \cdot [-2 \Delta_y (y_t - y_s) - 2 \Delta_x (x_t - x_s)] +    $ 
$\alpha   \cdot [  \Delta_x (y_t - y_s) -   \Delta_y (x_t - x_s)] = 0  $.\\  

Finally:
\begin{equation}
 \alpha = \frac{m_2 \tau^2 + m_1 \tau  + m_0}{n_2 \tau^2 + n_1 \tau + n_0} ,
\end{equation}

where \\ \\
\begin{minipage}{0.45\linewidth}
 $m_2 =    \Delta_y (x_0 - x_s) -   \Delta_x (y_0 - y_s) + k $,\\
 $m_1 =  - 2 \Delta_x (x_0 - x_s) - 2 \Delta_y (y_0 - y_s) $, \\
 $m_0 =  -m_2 +2k                     $, \\
\end{minipage}
\hfill
\begin{minipage}{0.45\linewidth}
$n_2 =    \Delta_y (x_t - x_s) -   \Delta_x (y_t - y_s) $, \\
$n_1 =  - 2 \Delta_x (x_t - x_s) - 2 \Delta_y (y_t - y_s) $, \\
$n_0 =  -n_2                       $ and \\
\end{minipage}

$\Delta_x = (x_2 - x_1)$ , $\Delta_y = (y_2 - y_1)$ , $k = x_1 y_2 - x_2 y_1 $. 

\end{document}